\documentclass[11pt]{article}
\usepackage{amsmath}
\usepackage{amssymb}
\usepackage[dvips]{epsfig}

 \advance \textwidth by -2in
 \advance \textheight by -3in
\def\##1{{\underline #1}}
\def\*#1{\tilde{{\underline #1}}}
\def\=#1{\underline{\underline{#1}}}
\def\+#1{\tilde{\underline{\underline{#1}}}}

\def\le{\left(}
\def\ri{\right)}

\def\c#1{\cite{#1}}

\def\r#1{(\ref{#1})}

\def\.{\mbox{ \tiny{$^\bullet$} }}
\def\,{\thinspace}

\def\rt{\le \#x,t\ri}
\def\ro{\le \#x,\omega\ri}

\def\curl{\nabla\times}
\def\div{\nabla\.}

\begin{document}

\noindent{\it {\large On electromagnetic boundary conditions and constitution of homogeneous materials}}
\vskip 0.2cm

\noindent  {Akhlesh Lakhtakia}\footnote{Tel: +1 814 863 4319; Fax: +1 814 865 9974;
E--mail: akhlesh@psu.edu}
\vskip 0.2cm
\noindent {\em Computational \& Theoretical Materials Sciences Group (CATMAS)\\
Department of Engineering Science \& Mechanics\\
Pennsylvania State University, University Park, PA 16802--6812, USA}\\
\medskip
and\\
\medskip
\noindent{\em Department of Physics, Imperial College, London SW7 2AZ, United Kingdom}

\noindent {\bf ABSTRACT:} Constitutive scalars that are absent in the fundamental
differential equations of frequency--domain
electromagnetism on the two sides of a boundary between
two homogeneous mediums are shown to disappear from the boundary
conditions as well.

\section{Introduction}
Modern macroscopic electromagnetism employs four differential equations containing four vector
fields, a source charge density, and a source current density. Two of the four fields
 (i.e., $\*E\rt$ and $\*B\rt$) are primitive fields, while the remaining two (i.e., $\*D\rt$ and $\*H\rt$)
 are induction fields. Unlike the primitive fields, the induction fields do not have microscopic existence and are
 not considered fundamental but undoubtedly offer a great deal of convenience
 \c{vanBladel,W03}.
 
 From the four differential equations emerge four boundary conditions or jump
 conditions, by employing  well--known  constructions involving pillboxes and
 contours \c{DJ,Kraus}. 
 This communication arose from an examination of the piecewise uniqueness of the
 induction field phasors and the consequences for boundary conditions 
 in macroscopic frequency--domain electromagnetism. It provides a
 negative answer to the following question:  Can
 constitutive scalars that are absent in the fundamental
differential equations  on the two sides of a boundary between
two homogeneous mediums appear in the boundary
conditions?
 
 \section{The question}
Let us consider all space to be divided into two distinct regions, $V_{+}$ and
$V_{-}$, separated by a boundary $S$. The sources of the electromagnetic field
are located in $V_+$ sufficiently far away from $S$, whereas our analysis concerns
the immediate vicinity of $S$ on both sides. Both regions are filled with
different spatially homogeneous and temporally invariant   mediums.

The frequency--domain  Maxwell equations
\begin{equation}
\left.\begin{array}{l}
\div\#B\ro=0\\[5pt]
\curl\#E\ro -
i\omega{\#B}\ro = \#0
\end{array}\right\}\,,\quad \#x\in V_+\cup V_-,
\label{eqs1}
\end{equation}
are applicable in both $V_+$ and $V_-$, with $i=\sqrt{-1}$, $\omega$ as the angular
frequency, and
 $\#B\ro$ as the temporal Fourier transform
of $\*B\rt$, etc. 
The
remaining two frequency--domain  Maxwell equations in $V_+$ may be written as
\begin{equation}
\left.\begin{array}{l}
\div\#D\ro=\rho_{so}\ro\\[5pt]
\curl{\#H}\ro +i\omega{\#D}\ro = 
{\#J}_{so}\ro
\end{array}\right\}\,,\quad \#x\in V_+\,,
\label{eqs2+}
\end{equation}
where $\rho_{so}$ and $\#J_{so}\ro$ are the  source densities. In the region
$V_-$, the corresponding equations are as follows:
\begin{equation}
\left.\begin{array}{l}
\div\#D\ro=0\\[5pt]
\curl{\#H}\ro +i\omega {\#D}\ro = 
\#0
\end{array}\right\}\,,\quad \#x\in V_-\,.
\label{eqs2-}
\end{equation}

As may be gleaned from textbooks \c{Jack}, the boundary conditions 
\begin{equation}
\left.\begin{array}{ll}
\#B^{norm}({\#x}+,\omega) =\#B^{norm}({\#x}-,\omega) \\[5pt]
\#E^{tan}({\#x}+,\omega) = \#E^{tan}({\#x}-,\omega)\\[5pt]
\#D^{norm}({\#x}+,\omega) =\#D^{norm}({\#x}-,\omega) \\[5pt]
\#H^{tan}({\#x}+,\omega) = \#H^{tan}({\#x}-,\omega)
\end{array}\right\}\,, \quad \#x\in S\,,
\label{eqs3}
\end{equation}
are derivable from \r{eqs1}--\r{eqs2-}.  Here,  $\#B^{norm}({\#x}\pm,\omega)$
indicate the normal components of $\#B\ro$ on the sides of $S$ in $V_\pm$, 
whereas  $\#E^{tan}({\#x}\pm,\omega)$
denote the tangential components of $\#E\ro$ similarly, etc. The boundary
conditions \r{eqs3}$_1$ and \r{eqs3}$_2$ involve the primitive field phasors,
whereas \r{eqs3}$_3$ and \r{eqs3}$_4$ involve the induction field phasors.
We have also taken note of the location of the source densities chosen as not being in
the vicinity of $S$.

Let us decompose the induction field phasors as follows:
\begin{equation}
\left.
\begin{array}{ll}
\#D\ro = \#D_a\ro +\alpha_\pm(\omega)\,\#B\ro\\[5pt]
\hspace{2cm}-\beta_\pm(\omega)\,\curl\#A\ro\,\\[5pt]
\#H\ro = \#H_a\ro -\alpha_\pm(\omega)\,\#E\ro\\[5pt]
\hspace{2cm}+i\omega\beta_\pm(\omega)\,\#A\ro
+\gamma_\pm(\omega)\,\nabla\phi\ro\,
\end{array}\right\}
\quad \#x\in V_\pm\,.
\label{eqs4}
\end{equation}
The terms on the right sides of \r{eqs4}$_1$ and \r{eqs4}$_2$
are mutually exclusive.
Whereas  $\#A\ro$ and $\phi\ro$ are some fields that are not necessarily
electromagnetic, the six scalars $\alpha_\pm(\omega)$, etc.,
are uniform in the respective regions and may be considered as
{\em  constitutive
scalars\/}.

A blind application of the boundary conditions \r{eqs3}$_3$ and \r{eqs3}$_4$
would involve all six constitutive scalars appearing on the right sides of \r{eqs4}$_1$
and \r{eqs4}$_2$. Would that be correct?

\section{The answer}
Boundary conditions in electromagnetism are not externally imposed
but instead emerge from the fundamental differential equations \c{JZB85}. Therefore,
in order to answer the question posed in the previous section, let us first substitute \r{eqs4} in
\r{eqs2+} and \r{eqs2-}. The resulting differential equations are as follows:
\begin{equation}
\left.\begin{array}{l}
\div\#D_a\ro=\rho_{so}\ro\\[5pt]
\curl{\#H}_a\ro +i\omega{\#D}_a\ro = 
{\#J}_{so}\ro
\end{array}\right\}\,,\quad \#x\in V_+\,,
\label{eqs5+}
\end{equation}
\begin{equation}
\left.\begin{array}{l}
\div\#D_a\ro=0\\[5pt]
\curl{\#H}_a\ro +i\omega {\#D}_a\ro = 
\#0
\end{array}\right\}\,,\quad \#x\in V_-\,.
\label{eqs5-}
\end{equation}
The  following {\em correct} boundary conditions emerge from the application
of the standard techniques \c{DJ,Kraus} to \r{eqs1}, \r{eqs5+}, and \r{eqs5-}:
\begin{equation}
\left.\begin{array}{ll}
\#B^{norm}({\#x}+,\omega) =\#B^{norm}({\#x}-,\omega) \\[5pt]
\#E^{tan}({\#x}+,\omega) = \#E^{tan}({\#x}-,\omega)\\[5pt]
\#D^{norm}_a({\#x}+,\omega) =\#D^{norm}_a({\#x}-,\omega) \\[5pt]
\#H^{tan}_a({\#x}+,\omega) = \#H^{tan}_a({\#x}-,\omega)
\end{array}\right\}\,, \quad \#x\in S\,.
\label{eqs6}
\end{equation}
Let us note that these boundary conditions do not involve the six 
constitutive scalars
$\alpha_\pm(\omega)$, $\beta_\pm(\omega)$, and $\gamma_\pm(\omega)$,
which are also absent from the fundamental equations  \r{eqs1}, \r{eqs5+}, and \r{eqs5-}.
These constitutive scalars can therefore be safely set to zero without affecting macroscopic electromagnetism.

\section{Concluding remarks}
The foregoing exercise
shows that constitutive scalars that are absent in the fundamental
differential equations of frequency--domain electromagnetism on the two sides of a boundary between
two homogeneous mediums cannot appear in the boundary
conditions. If wrongly allowed to appear by virtue of
a blind application of the standard boundary conditions, these
constitutive scalars would lead to wrong conclusions from the
solutions of boundary value problems.

In other words, the Maxwell equations act like filters of constitutive terms, and
a constitutive terms filtered out of the Maxwell equations is also filtered out of
the boundary conditions.

\end{document}